\def\year{2022}\relax
\documentclass[letterpaper]{article} 
\usepackage{aaai22}  
\usepackage{times}  
\usepackage{helvet} 
\usepackage{courier}  
\usepackage[hyphens]{url}  
\usepackage{graphicx} 
\usepackage{natbib}  
\usepackage{caption} 
\usepackage{booktabs} %
\usepackage{multirow}
\usepackage{graphicx}
\usepackage{enumerate}
\usepackage{subfigure}
\usepackage{threeparttable}
\usepackage{algorithmicx}
\usepackage[ruled,linesnumbered]{algorithm2e}
\usepackage{algpseudocode} 
\usepackage{amsmath}
\usepackage{amssymb}
\usepackage{bbm}
\usepackage{bbding} %
\usepackage{comment}
\usepackage[title]{appendix}
\title{Self-supervised Representation Learning for Trip Recommendation}
\author{
    Qiang Gao\textsuperscript{$\dag$}\thanks{Corresponding Author(qianggao@swufe.edu.cn).}
    ,
    Wei Wang\textsuperscript{$\dag$},
    Kunpeng Zhang\textsuperscript{$\S$},
    Xin Yang\textsuperscript{$\dag$},
    Congcong Miao\textsuperscript{$\ddag$}
    \\
}
\affiliations{
    \textsuperscript{$\dag$}Southwestern University of Finance and Economics, Chengdu, China.\\
     \textsuperscript{$\S$}University of Maryland, College park.\\
    \textsuperscript{$\ddag$}Tsinghua University, Beijing, China.\\
    qianggao@swufe.edu.cn, wangwei1997@smail.swufe.edu.cn,
    kpzhang@umd.edu,
    yangxin@swufe.edu.cn,
    mccmiao@163.com
}

\begin{document}

\maketitle

\begin{abstract}
Trip recommendation is a significant and engaging location-based service that can help new tourists make more customized travel plans. It often attempts to suggest a sequence of points of interest (POIs) for a user who requests a personalized travel demand. Conventional methods either leverage the heuristic algorithms (e.g., dynamic programming) or statistical analysis (e.g., Markov models) to search or rank a POI sequence. These procedures may fail to capture the diversity of human needs and transitional regularities. They even provide recommendations that deviate from tourists' real travel intention when the trip data is sparse. Although recent deep recursive models (e.g., RNN) are capable of alleviating these concerns, existing solutions hardly recognize the practical reality, such as the diversity of tourist demands, uncertainties in the trip generation, and the complex visiting preference. Inspired by the advance in deep learning, we introduce a novel self-supervised representation learning framework for trip recommendation -- \textbf{SelfTrip}, aiming at tackling the aforementioned challenges. Specifically, we propose a two-step contrastive learning mechanism concerning the POI representation, as well as trip representation. Furthermore, we present four trip augmentation methods to capture the visiting uncertainties in trip planning. We evaluate our SelfTrip on four real-world datasets, and extensive results demonstrate the promising gain compared with several cutting-edge benchmarks, e.g., up to 4\% and 10\% improvements on Osaka regarding $F_1$ and pair-$F_1$.
\end{abstract}

\section{Introduction}
\label{Introduction}
The ubiquitousness of GPS-integrated mobile devices and the wide use of location-based services have enabled the generation of massive volumes of geo-tagged data, which offers unprecedented opportunities to explore human moving intentions and correspondingly conduct various downstream location-based applications, e.g. POI recommendation~\cite{yu2020category}, friend recommendation~\cite{wu2019graph}, trajectory classification~\cite{Gao2017}, and trip recommendation~\cite{he2019joint}, to name a few. Especially, trip recommendation (a.k.a, itinerary recommendation) has drawn widespread attention from researchers and practitioners in recent years, aiming at suggesting a sequence of point-of-interests (POIs) while meeting a set of user-provided constraints (e.g., starting and ending points, the number of POIs to be visited). Trip recommendation is significantly different from traditional route planning in that the latter is usually to search for the shortest or a minimum cost-based route. In contrast, trip recommendation takes more into consideration the diversity and personalized needs of a POI sequence, which is also not constrained by the road network~\cite{wang2021personalized}. Therefore, trip recommendation is a more challenging task in location-based recommendation services.

A conventional solution for trip recommendation is to employ an orienteering-based method to maximize the user-specific query constraints, which is inspired by the navigation operation. Specifically it mines and incorporates prior knowledge extracted from historical trips (e.g., the attractiveness of POIs and POI categories) to offer a sequence of POIs via POI generation~\cite{Lim2015,chen2016learning,taylor2018travel}. For example, ~\cite{he2019joint} uses an integer linear programming to explore the exact optimal trips. However, heuristic solutions usually either rely on the local transitional distribution between POIs or only pursue the shortest time budget, leading to the failure in respect of estimating the conditional distribution of human transition regularity, considering the diversity of human demands, and exploiting the high-ordered relations. Complement to these, human mobility actually contains many explicit and implicit characteristics, e.g, the semantic proximity of POIs and the spatial-temporal dependencies. In this aspect, another solution was to leverage deep representation learning methods to explore the nature of human characteristics, e.g, context-aware POI embedding~\cite{he2019joint,ho2021user} and recursive trip modeling~\cite{zhou2020semi,ho2021user,gao2021adversarial}. Under real datasets, they have empirically demonstrated superior performance over traditional methods, which inspires us to tap the charm of deep representation learning to be competent for trip recommendation. 

However, trip recommendation is still facing several challenges, including: (1) \textit{The heterogeneity of user demands:} Due to the diversity of travel plans requested by users, it is impossible for us to perceive or grasp the needs of all users. The reality is often that we only observe partial query records and their corresponding trips. (2) \textit{The uncertainty in trips:} Although we can provide tourists with a definite travel route according to a given query, it is difficult for us to understand user’s underlying true travel preference based on a small amount of available trip data, as well as the heterogeneity of individual preferences, which brings the risk of uncertainty during trip recommendation. (3) \textit{The complex visiting patterns:} Due to the data sparsity issue, it is intricate to explore the higher-ordered transitional regularities, especially the long-term dependencies.

Recent self-supervised learning techniques, e.g., contrastive learning, have achieved considerable success for deep representation learning in natural language processing, computer vision, and recommender systems~\cite{liu2021self}. They are capable of discovering implicit supervised signals together with multiple versions of data augmentation strategies for promoting representation learning. Motivated by this, we propose a novel self-supervised framework, namely SelfTrip (\textbf{Self}-supervised Representation Learning for \textbf{Trip} Recommendation), to remedy the aforementioned concerns. In our SelfTrip, we design a novel POI representation learning module based on contrastive learning, where a causal random walk strategy is proposed to enrich the diversity of query demands. To alleviate the uncertainties in trip data, we leverage four trip data augmentation methods to stimulate human travel behaviors, whereafter a contrastive trip presentation learning module is proposed. Finally, we model the trip data with a conditional (query-based) recursive module for trip generation, where a destination-enabled supervised signal regarding the trip destination is incorporated to constrain POI generation bias. In summary, our work make the following contributions:
\begin{itemize}
    \item To the best of our knowledge, SelfTrip is among the first framework to respectively improve the representation of POI and human mobility with two contrastive learning schemes.
    \item We investigate the complex interactions between queries and potentially visiting POIs on a huge query-based POI sequence set. Accordingly, the self-supervised POI learning is proposed to formulate the context-aware POI representation.
    \item We design four data augmentation strategies to mimic human travel behaviors based on the sparse trip data. In the sequel, we present a self-supervised trip learning to enhance the trip inference, where we also incorporate a destination-enabled supervised signal to constrain the POI prediction.
    \item Experimental results conducted on four real-world datasets demonstrate that our SelfTrip significantly outperforms several cutting-edge benchmarks.
\end{itemize}

\section{Preliminaries}
\label{Preliminaries}
In this section, we introduce basic definitions and the problem formulation. We also present the contrastive loss used later in our SelfTrip.

\textbf{Definition 1: Trip.}
Let $L$ denote a set of POIs and $\mathcal{T}$ represent a set of historical trips. Without loss of generality, we use a triplet $l: <id,lo,la>$ to represent a POI $l\in L$, where $id$, $lo$ and $la$ represent the POI id, longitude and latitude, respectively. Formally, a trip is a POI sequence ordered by check-in time, e.g., $<l_1,t_1>\to<l_2,t_2>\to\cdots\to<l_m,t_m>$, where $<l_*,t_*>$ denotes a POI $l_*$ in a trip
was visited at time $t_*$.

\textbf{Definition 2: Query.}
In this study, a query $q$ is a quintuple $<l_s,t_s,l_d,t_d,N>$ that consists of the source POI $l_s$, starting time $t_s$, the destination POI $l_d$, ending time $t_d$, and the number of POIs $N$ to visit.

\textbf{Trip Recommendation.}
  Given historical trips $\mathcal{T}$, the trip recommendation system can return a trip $T=(l_1,l_2,\cdots,l_N)$ if a tourist provides a query $q:<l_s,t_s,l_d,t_d,N>$, where $l_s=l_1$ and $l_d=l_N$. It can be formally described as follows:
\begin{equation}
    \label{TR}
    \tilde{T}=\arg\max_\theta \mathcal{P}(T|q,\mathcal{T}).
\end{equation}

\textbf{Contrastive Loss.}
As a widely used technique in self-supervised learning, contrastive learning aims to construct positive samples and negative samples based on the original data, which makes similar samples closer in the projection space while dissimilar samples are relatively further away. Among which Noise Contrastive Estimation (NCE) becomes very popular in various applications~\cite{oord2018representation,yan2021consert,aitchison2021infonce}. Its overall objective is defined as:
\begin{small}
\begin{equation}
\label{nce}
\mathcal{J}_{\mathrm{NCE}}=\mathbb{E}\left[\log \left(\frac{e^{g(x)^{\top} g\left(x^{+}\right)}}{e^{g(x)^{\top} \operatorname{g}\left(x^{+}\right)}+\sum_{j=1}^{J} e^{g(x)^{\top} g\left(x_{j}^{-}\right)}}\right)\right],
\end{equation}
\end{small}
where $x$ refers to the `anchor', $x^+$ and $x^-$ respectively represent the positive and negative sample. $g$ and $J$ respectively denote the learning function (e.g., deep neural networks) and the number of negative samples.

\section{Methodology}
\label{Methodology}
We turn to introduce our proposed SelfTrip. First, we present an overview of the architecture design, whereafter elaborating the POI learning and trip learning procedures. Finally, we describe the training details of SelfTrip.
\begin{figure}[ht]
    \vspace{-0.4cm}
    \centering
    \includegraphics[width=0.48\textwidth]{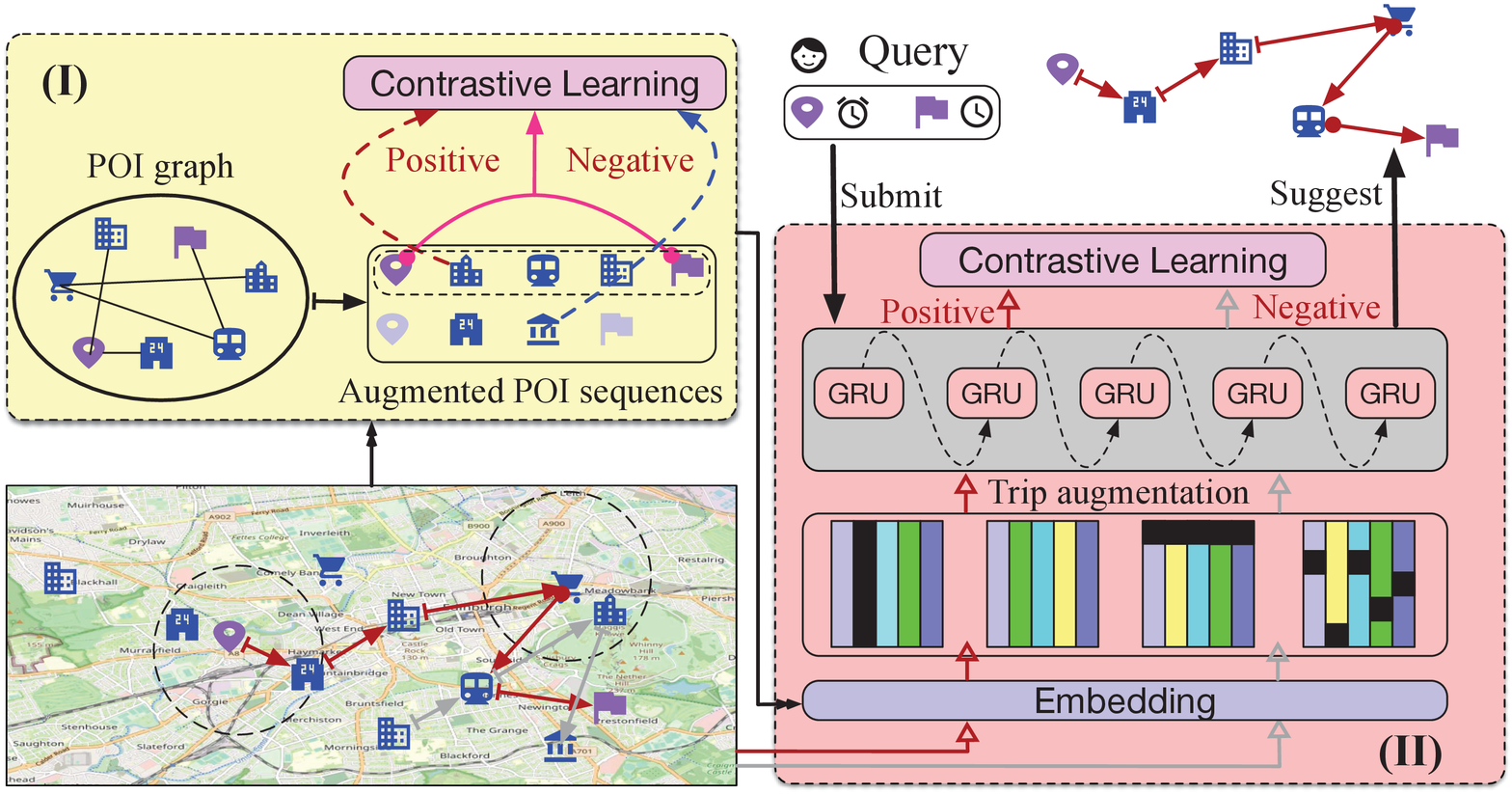}
    \caption{The overview of the SelfTrip framework.}
    \label{framew}
    \vspace{-0.4cm}
\end{figure}
\subsection{Architecture}
As Fig.\ref{framew} shows, SelfTrip mainly contains two components: \textbf{(I) Self-supervised POI Learning} aims to embed each POI into a low-dimensional space. This is an essential action for trip recommendation, which not only mitigates the problem of the curse of dimensionality but also is capable of exploring the semantic relationships between queries and POIs. \textbf{(II) Self-supervised Trip Learning} first formulates a context-aware query vector, and then randomly selects designed trip augmentation methods to generate variants of an original trip. Next we present a query-based GRU (Gated Recurrent Unit) as the basic encoder for contrastive learning. Besides, we use this encoder as the trip generator in the formal training process.

\subsection{Self-supervised POI Learning}
The intuitive idea of trip recommendation is to use a query provided by a user to search out a sequence of POIs, thus we expect to design such an embedding module that can preserve interactions between the query and POIs as much as possible. However, it is not easy to full capture their interactions due to the scale of the training dataset and the sparsity of query records. To this end, we first construct a POI graph to describe the interactions among POIs. Then we adopt a simple interaction augmentation strategy to integrate potential geographical preferences of tourists. To obtain more possible queries, we propose a novel query-based sampling approach to produce more reasonable trips from the constructed POI graph. In the end, we present a contrastive POI learning method to learn possible interactions between queries and POIs.

\subsubsection{Augmented POI graph.} To extract query-POI interactions, we transform historical trips into a basic POI graph $G_a(L,E)$, where $E$ denotes a set of edges reflecting the successive visiting behaviors. To incorporate potential geographical preferences of tourists, we collect each POI’s geographical neighbors by setting a distance threshold, and correspondingly building an edge between this POI and
all of its geographical neighbors that are within this threshold. To constrain the scale of neighbors, we set the distance
threshold to 3 km. Finally, we obtain an augmented POI graph $G_a(L,E_a)$, where $E \subsetneqq E_a$. 
\subsubsection{Causal Random Walks.}
Before we sample trips from $G_a$, we create a transition matrix $A$ to describe the interaction probability between pairs of POIs (we actually refer to such two POIs as the source POI and destination POI) according to $G_a$. More specifically, given any two nodes $l_i$ and $l_j, (l_i \in L, l_j \in L)$, the transition probability from $l_i$ to $l_j$ is calculated by:
\begin{equation}
    P(l_j|l_i)=f_{ij}/f_i.
\end{equation}
Where $f_{ij}$ denotes the frequency of edge $l_i \to l_j$ appeared in $E_a$, and $f_i$ is the frequency of $l_i$ appeared in $E_a$. Thus, we finally obtain the transition matrix $A$ corresponding to $G_a(L,E_a)$. Next, we use the causal random walk strategy to generate a set of POI sequences based on $G_a$ and $A$. There exist $L$ POIs in $\mathcal{T}$, thereby we traverse all unique POI pairs (e.g., $<l_i,l_j>$) from the graph $G_a$ and obtain a set of query candidates $\mathcal{Q}=\{<l_i,l_j>|i\neq j,l_i\in L,l_j\in L\}$. Actually, there exist at most $|\mathcal{T}|$ queries when and only when the query of each trip in $\mathcal{T}$ is different, meanwhile, we usually confront $|\mathcal{T}|\ll |\mathcal{Q}|$ due to the shortage of training data. Thus, we use $G_a$ and $A$ to generate $M$ POI sequences corresponding to each query candidate in $\mathcal{Q}$, which enables to alleviate the lack of query records in given trip data.

More specifically, given a query $<l_i,l_j> (<l_i,l_j> \in \mathcal{Q})$, we simulate a random walk with a length budget $\alpha$. Let $\ell_\kappa=(\kappa\ge 1)$ be the $\kappa$th node in a walk, where we set $\ell_0$=$l_i$. Correspondingly, node $\ell_\kappa$ is generated by the distribution $P(\ell_\kappa|\ell_{\kappa-1})$, where $P(\ell_\kappa|\ell_{\kappa-1}) \in A$. Once $\ell_\kappa=l_j$ and $\kappa \le \alpha$, we store the generated POI sequence $\{\ell_0, \ell_1,\cdots,\ell_\kappa\}$ and proceed to the next walk. If $\ell_1=l_j$ or $\kappa \ge \alpha$, we will drop the current POI sequence and proceed to the next walk directly. The reason we set a length budget $\alpha$ is that people cannot have enough time to take a long trip or it is unrealistic to generate a long trip for the tourist to visit. Due to the constraint of $\alpha$, the generated POI sequences will have different lengths, which, to some extent, allows us to cater to the real scenarios of visiting behaviors. In the end, we will collect a larger set of POI sequence $\mathcal{S}$, and use it as the training set for the following POI representation learning.

\subsubsection{Contrative POI Learning.}
We now use the contrastive learning to train the POI representation for capturing the potential interactions between queries and POIs. And we expect to maximize the similarity between a given query and the POI sampled in the same POI sequence while minimizing the similarity between a given query and the POI sampled from a different sequence. Given a POI sequence $S'=\{l_1,l_2,\cdots,l_N\}$ sampled from $\mathcal{S}$, we set up a learnable matrix $\mathbf{v}\in \mathbb{R}^{|L|\times d}$ as the initial POI embeddings, where $d$ denotes the embedding size. Next, we set $l_1$ as the source POI and $l_N$ as the destination POI, and using a simple averaging operation to formulate a query representation $\mathbf{q}_{S'}$, where $\mathbf{q}_{S'}=\frac{1}{2}(\mathbf{v}'(l_1)+\mathbf{v}'(l_N))$ ($\mathbf{v}' \in \mathbb{R}^{|L|\times d}$ refers to the leanable matrix corresponding to user query). We regard $\mathbf{q}_{S'}$ as the `\textit{anchor}' and randomly select a POI $l$ in ${S'}$ as the positive sample while randomly select a POI $l'$ from another 
sequence as the negative sample in which it is associated with a query that is distinct from $\mathbf{q}_{S'}$. For simplicity, let $\mathbf{q}_{S'}$ denote the original pair $(l_1,l_N)$. Recall Eq.(\ref{nce}), we randomly sample $k-1$ negative samples to cater to the training requirements of contrastive learning, i.e., $\{l'_j|l'_j \in S \setminus S'\}_1^{k-1}$. Therefore, the objective for trip $S'$ is:
\begin{small}
\begin{align}
\label{obj-poi}
&\mathcal{J}_{(\mathbf{q}_{S'})}=\\
&\mathbb{E}_{l \in S'\setminus q_{S'}}[s \left(\mathbf{v}(l), \mathbf{q}_{S'}\right)
-\log\sum_{j=1}^{k-1} \exp (s \left(\mathbf{v}(l'_j), \mathbf{q}_{S'}))\right],\nonumber
\end{align}
\end{small}
where `$\setminus$' defines subtraction operation of set and $s$ refers to the $cosine$ function. Finally, we use the optimal $\mathbf{v}$ as the POI embeddings.
\subsection{Self-supervised Trip Learning}
In this section, we first show how to formulate a dense query representation and trip representation. Then, we present four trip augmentation strategies for sparse trips and describe how to train our trip representation with contrastive learning. In the end, we introduce the training mechanism of SelfTrip.

\subsubsection{Query Encoder.} To accommodate the input of our SelfTrip, it is necessary to transform a given query $q:<l_s,t_s,l_d,t_d,N>$ into a couple of dense representations. Hence, we first encode the source POI and destination POI in the query into dense representations. Since we have trained the POI embeddings $\mathbf{v}$ from the above, we can directly use it to obtain the corresponding dense representations for $l_s$ and $l_d$, whereas they can be defined as $\mathbf{v}(l_s)$ and $\mathbf{v}(l_d)$. Inspired by previous works on time information processing~\cite{gao2021adversarial}, an hour-level representation strategy is adopted to encode $t_s$ and $t_d$, which splits the day into 24 discrete time intervals and uses simplest one-hot encoding method to represent the 24 hour-level intervals. For example, a tourist whose ending time $t_d$ can be represented as $\mathbf{u}(t_d) \in \mathbb{R}^{d'}$, where $d'$ is the embedding dimension. 

Traditional methods usually use a simple concatenation operation to mingle those dense representations into an unified vector to represent the given query $q$. To capture the interactions of $l_s$, $t_s$, $l_d$, and $t_d$, we design a transformed operation to obtain a final representation for query $q$:
\begin{align}
\label{query}
    \mathbf{q}=\text{LeakyRelu}([\mathbf{v}(l_d)\rVert \mathbf{u}(t_d)]\sum ([\mathbf{v}(l_s)\rVert \mathbf{u}(t_s)]K_q)\\\nonumber
    +[\mathbf{v}(l_s)\rVert \mathbf{u}(t_s)\rVert \mathbf{v}(l_d)\rVert \mathbf{u}(t_d)]W_q+b_q).
\end{align}
Where $\rVert$ denotes the concatenation operation, $W_q \in \mathbb{R}^{2(d+d')\times d''}$ is a learnable matrix, $b_q \in \mathbb{R}^{d''}$ is the bias, and $K_q \in \mathbb{R}^{(d+d')\times (d+d')\times d''}$ is a third-order tensor.

\textbf{Trip Encoder.}
We now introduce how to model a given query $q$ and its associated trip $T$ by a query-based recurrent neural network. That is to say, we aim to incorporate the query information into the trip representation in a recursive manner. We first obtain the context-aware $\mathbf{q}$ to represent the given $q$. For trip $T=\{l_1,l_2,\cdots,l_N\}$, we transform each discrete POI to dense representation by looking up the optimized $\mathbf{v}$, and obtain the basic trip representation $\mathbf{T}=\{\mathbf{v}(l_{1}),\mathbf{v}(l_{2}), \cdots, \mathbf{v}(l_{N})\}$. Next, we employ the query-based GRU to capture the dependencies among the POIs in $T$. Taking $l_\tau \in T$ as an example, its hidden state can be formulated as:
\begin{align}
\label{encoder}
\mathbf{h}_\tau=\text{GRU}([\mathbf{v}(l_{\tau})\rVert f(\mathbf{q})],\mathbf{h}_{\tau-1}),
\end{align}
where $f(\cdot)$ denotes a dense layer. In the end, we can obtain the hidden state of each POI, i.e., $\mathbf{h}_1,\cdots,\mathbf{h}_N$.

\textbf{Trip Augmentation.} As a prerequisite, trip representation with self-supervised learning needs to set multi-views of real training trips as the positive samples. Inspired by recent contrastive text generation~\cite{yan2021consert} and contrastive item augmentation~\cite{zhou2020s3}. We present four strategies for trip data augmentation which aims to mimic the human real visiting behaviors. They are:
\begin{itemize}
    \item \textit{POI Mask}. It is a simple but realistic strategy. For instance, the tourist Bob would choose some of POIs to visit although we provide him a longer trip. Thus, we randomly mask some POIs in a given POI sequence to generate a new trip.

    \item \textit{Shuffling}. Although we provide the tourist with a possible itinerary, he/she may change their visiting plan during travelling due to the personal interests. For instance, SelfTrip provides an itinerary $<A,B,C>$ for tourist Bob, when Bob prepares to visit $B$ after visiting $A$, he may decide to visit $C$ first because he considers $C$ more attractive to him or $C$ is an urgent need. To simulate the realistic human moving intention, we change the order of a given sequence of POIs to form a new trip and add it to our training set.
    
    \item \textit{Feature Cutoff}. Since we use the dense representation to tackle the discrete POI or trip, the feature cutoff attempts to erase some feature dimensions to generate an augmentation view of input data. To be specific, it only applies to the POI embeddings before we model trip learning.
    
    \item \textit{Dropout.} Dropout is a widely used approach to alleviate the over-fitting issue during neural network modeling~\cite{hinton2012improving}. Hence, we use it to randomly drop part of values out in each POI embedding by a certain probability (e.g., the dropout rate is 0.5) and set each corresponding value to zero.
\end{itemize}
To accommodate the following contrastive trip learning, we feed each trip $T$ in $\mathcal{T}$ into the augmentation layer and randomly select two strategies for augmentation. For example, using the augmentation layer to generate two views of $\mathbf{T}$, denoted as $\hat{\mathbf{T}}^p$ and $\hat{\mathbf{T}}^q$.

\textbf{Contrastive Trip Learning.}
Notably, using larger mini-batch is a common method for contrastive learning, we thus choose $m$ queries $\{q_i\}_{i=1}^{m}$ and their corresponding trips $\{T_i\}_{i=1}^{m}$ from the training set to satisfy the requirement of mini-batch. Next, we use trip augmentation layer to generate the augmented trip set $\{\hat{\mathbf{T}}_i^p\}_{i=1}^{m}$ and $\{\hat{\mathbf{T}}_i^q\}_{i=1}^{m}$, where each ${T}_i$ is associated with $\hat{\mathbf{T}}_i^p$ and $\hat{\mathbf{T}}_i^q$. Correspondingly, we can obtain the context-aware query set $\{\mathbf{q}_i\}_{i=1}^{m}$ by Eq.~(\ref{query}). As for augmented trips, we use the Trip Encoder to obtain the hidden states of these trips, and employ the final state of each augmented trip as input of contrastive learning. For instance, we respectively use $\mathbf{h}_{i}^p$ and $\mathbf{h}_{i}^q$ to represent the final states of $\hat{\mathbf{T}}_i^p$ and $\hat{\mathbf{T}}_i^q$, and use them to represent these two trips, denoted by $\tilde{\mathbf{T}}_i^p$ and $\tilde{\mathbf{T}}_i^q$ for consistency. Similarly, we can formulate the dense representations of other augmented trips, denoted by $\{\tilde{\mathbf{T}}_i^p\}_{i=1}^m$ and $\{\tilde{\mathbf{T}}_i^q\}_{i=1}^m$. Now, we can treat any $<\tilde{\mathbf{T}}_i^p$,$\tilde{\mathbf{T}}_i^q>$ as the positive pair, while regarding the remain $(m-1)$ pairs as the negative. Thus, the objective regarding $T_i$ can be defined as:
\begin{small}
\begin{equation}
    \label{obj-trip}
    \mathcal{J}_{(i)}=\log \frac{\exp(s(\tilde{\mathbf{T}}_i^p,\tilde{\mathbf{T}}_i^q))}{\exp(s(\tilde{\mathbf{T}}_i^p,\tilde{\mathbf{T}}_i^q))+\sum_{j=1,j \neq i}^{m} \exp(s(\tilde{\mathbf{T}}_i^p,\tilde{\mathbf{T}}_j^q))}
\end{equation}
\end{small}
In practical implementation, we can train the trips $\{T_i\}_{i=1}^{m}$ in a parallel manner by: 
\begin{small} 
\begin{align}
\label{con-trip}
&\mathcal{L}_{\{T_i\}_{1}^{m}}=
-\sum \mathbf{I} \odot \log  \text{softmax}(\left(\begin{array}{llll}
\tilde{\mathbf{T}}_1^p\\
\cdots\\
\tilde{\mathbf{T}}_i^p\\
\cdots\\
\tilde{\mathbf{T}}_m^{p}
\end{array}\right)^{\top}\left(\begin{array}{llll}
\tilde{\mathbf{T}}_1^q\\
\cdots\\
\tilde{\mathbf{T}}_i^q\\
\cdots\\
\tilde{\mathbf{T}}_m^{q}
\end{array}\right))
\end{align}
\end{small}
where $\mathbf{I}$ is the identity matrix (denoting the elements in diagonal are positive scores while others are negative scores) and $\odot$ denotes the element-wise Hadamard product. In the end, the overall loss function is:
\begin{equation}
\mathcal{L}_{self}=-\sum_{i=1}^{m \times |\mathcal{T}|} \mathcal{J}_{(i)}
\end{equation}
In the sequel, we will use the warmed-up Query Encoder and Trip Encoder for the following formal training.
\begin{figure}[ht]
    \centering
    \includegraphics[width=0.45\textwidth,height=0.12\textheight]{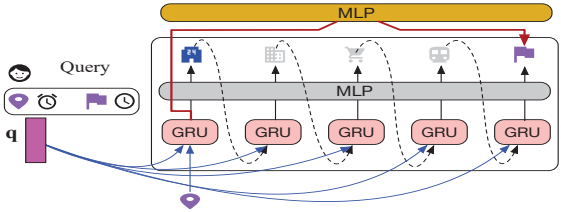}
    \caption{Trip generation with destination-enabled signals}
    \label{train}%
    \vspace{-0.5cm}
\end{figure}

\subsection{Formal Training}
We have used the self-supervised learning to embed each discrete POI to a low-dimensional vector, where the correlations among the query and POIs are incorporated. And we also used the self-supervised learning to pre-train model with contrastive loss. Now we turn towards using the warmed-up model for supervised trip training. Recall the objective of Eq.(1), SelfTrip is to generate the trip in a recursive manner as follows:
\begin{equation}
\tilde{l_\tau}=\text{GRU}([\mathbf{v}(l_{\tau-1})\rVert f(\mathbf{q})],\mathbf{h}_{\tau-1})W_f+b_f,
\end{equation}
where $W_f$ and $b_f$ are the learnable parameters.
Traditionally, we can form the final objective as:
\begin{equation}
\label{obj}
\mathcal{L}_{sup}(\theta)=-\sum_{i=1}^{|\mathcal{T}|} \log p\left(T_i \mid q\right),
\end{equation}
where $\theta$ refers to the parameters in SelfTrip.
However, trip recommendation is significantly different from traditional trajectory prediction tasks in that the latter works in a multi-round next POI generation manner~\cite{zhou2019context}. Actually, we usually neglect a supervised signal that each predicted POI is also constrained by the destination. Although such a constraint or interaction is unknown, we attempt to use a simple neural network (i.e., 
fully connected neural network) to stimulate such unknown interaction between the predicted POI and the destination. As Fig.~\ref{train} shows, we involve a destination-oriented signal to constrain the trip generation during the model training process. For a given query $q:<l_s,t_s,l_d,t_d,N>$, we assume that we have obtained a current hidden state $\mathbf{h}_\tau$. Then we use a one-layer fully connected neural network to build the relationship between $\mathbf{h}_\tau$ and the destination by: 
\begin{equation}
\tilde{l}_d= \text{softmax}\left(\boldsymbol{h}_{\tau} W'+b'\right),
\end{equation}
where $W'$ and $b'$ refer to the learnable parameters. Thus, we rewrite the Eq.(\ref{obj}) as:
\begin{align}
\label{obj-final}
&\mathcal{L}(\theta,\gamma)=\mathcal{L}_{sup}(\theta)+\mathcal{L}_{dest}(\theta,\gamma)= \\\nonumber
&
\sum_{1}^{|\mathcal{T}|} \sum_{\tau=1}^{N}( \log p(l_\tau \mid \mathbf{h}_{\tau-1},q)+\log p' (l_d \mid \mathbf{h}_{\tau-1},q )),
\end{align}
where $\gamma$ refers to  $W'$ and $b'$. Notably, the length $N$ in query $q$ will be considered as recursive rounds.

\section{Evaluation}
\label{Experiments}
We conduct experiments on four real-world datasets to demonstrate the effectiveness of our SelfTrip. In detail, we first introduce the datasets, benchmarking algorithms and evaluation metrics, followed by the model settings. Then, we present our results including overall performance comparison, the performance of individual module, study of query and trip augmentation, and parameter sensitivity.
\subsection{Dataset, Benchmark and Metrics}
\subsubsection{Datasets.}
As Table~\ref{dataset-label} shows, we use four real-world trip datasets, including Toronto, Osaka, Glasgow, and Edinburgh, from Flickr as our experimental data, which have been widely used in prior studies~\cite{chen2016learning,he2019joint,gao2021adversarial}.  To improve the quality of data, we do some regular data prepossessing, including filtering out short trajectories which contain less than three POIs, normalizing the timestamp into hour-level (i.e., mapping each timestamp into 24 intervals). Following~\cite{chen2016learning,he2019joint,gao2021adversarial}, we also adopt leave-one-out cross validation to evaluate all methods. 
\renewcommand{\arraystretch}{0.7} 
\begin{table}[ht]
\vspace{-0.2cm}
	\centering
	\setlength{\abovecaptionskip}{0pt}
	\setlength{\belowcaptionskip}{0pt}
	\caption{Descriptive statistics of datasets}
        \label{dataset-label}
	\fontsize{7.5}{8}\selectfont 
	\begin{tabular}{l|c|c|c}
		\toprule[1.1pt]
		\textbf{City}& \textbf{$\Phi$(Visits)} & \textbf{$\Phi$(Trajectory)}& \textbf{$\Phi$(User)}\\
		\hline
		$ \textbf{Edinburgh} $&33,944&5,028&1,454\\
	    $\textbf{Glasgow} $&11,434&2,227&601\\
		$\textbf{Osaka}$&7,747&1,115&450\\
		$\textbf{Toronto}$&39,419&6,057&1,395\\
		\bottomrule[1.1pt]
	\end{tabular}
	\vspace{-0.4cm}
\end{table}
\subsubsection{Benchmarks.}
We compare our SelfTrip with several recently developed representation learning-based methods, including:
\begin{itemize}
    \item \textbf{Markov}~\cite{chen2016learning}: It is a common and intuitive method where a POI transition matrix is constructed to recommend a trajectory.
    \item \textbf{POIRank}~\cite{chen2016learning}: It recommends a trajectory by first ranking POIs via a RankSVM method, and then connecting them based on their ranking scores.
    \item \textbf{Markov-Rank}~\cite{chen2016learning}: It constructs a POI transition matrix, and recommends a trajectory based on Markov transitions and POI ranking.
    \item \textbf{CATHI}~\cite{zhou2019context}: It is a recurrent neural network-based trajectory planning method in an end-to-end manner. We leverage an encoder to model the given query and use the decoder for trip recommendation.
    \item \textbf{C-ILP}~\cite{he2019joint}: It uses word2vec-based method to learn context-aware POI representation, followed by integer linear programming (ILP) for trip generation.
    \item  \textbf{DeepTrip}~\cite{gao2021adversarial}: It is an end-to-end trip recommendation method, while using a GAN-style neural network to approximate the similarity between the query and the trip.
    \item \textbf{NASR+}~\cite{wang2021personalized}: It is a deep neural network-based A* algorithm for route planning constrained by the road network.
    Since the check-in data is not constrained by the road network, we only use its RNN module enhanced by the attention mechanism for a fair comparison.
\end{itemize}

\subsubsection{Metrics.}
For performance comparison among all methods, we follow prior studies~\cite{Lim2015,he2019joint} to use two well-established metrics: (i) $F_1$ score to measure the quality of a recommended sequence -- which is, the harmonic mean of Precision and Recall of elements in a sequence; (ii) pairs-${F_1}$ score~\cite{chen2016learning,gao2021adversarial}, considers both POI correctness and sequential order by calculating the $F_1$ score of every pair of POIs, whether they are adjacent or not in a trajectory. Notably, the values of both pairs-${F_1}$ and ${F_1}$ are between 0 and 1. The higher the value, the better the recommended results, e.g., a value of 1 means that both POIs and their visiting order in the recommended trajectory are exactly the same as the ground truth. 

\subsubsection{Experimental Settings.}
We reproduce the benchmarks and implement our SelfTrip in Python while deep learning methods are accelerated by one NIVDIA RTX 3090 GPU. As for our SelfTrip, length budget $\alpha$ is 6, the batch size is 8, the dimensionality of POI is 250, the initial learning rate is 0.1, the hidden size in contrastive trip representation is 256, and the optimizer is Adam~\cite{kingma2014adam}. 
\renewcommand{\arraystretch}{0.7} 
\begin{table}[htp]
	\renewcommand{\multirowsetup}
    {\centering}
	\setlength{\abovecaptionskip}{0pt}
	\setlength{\belowcaptionskip}{0pt}
\caption{Overall performance comparison.}
\label{E-T}
\centering
\scriptsize
\setlength{\tabcolsep}{0.8mm}{
\begin{tabular}{l|cc|cc|cc|cc}
	\bottomrule[1pt]
	\multicolumn{1}{l|}{\multirow{2}{*}{\textbf{Method}}}&  
	\multicolumn{2}{c|}{\textbf{Edinburgh}}&\multicolumn{2}{c}{\textbf{Glasgow}}&	\multicolumn{2}{|c|}{\textbf{Osaka}}&\multicolumn{2}{c}{\textbf{Toronto}}\cr
	\cmidrule(lr){2-9}
	\textbf{}  &  $F_1$ & pairs-$F_1$& $F_1$ & pairs-$F_1$& $F_1$ & pairs-$F_1$& $F_1$ & pairs-$F_1$
	\\
	\hline
	Markov  & 0.645 &0.417  &0.725&0.495 &0.697&0.445 &0.669&0.407  \\
	POIRank& 0.700&0.432 &0.768&0.548  &0.745&0.511 &0.754&0.518\\
	Markov-Rank  &0.659&0.444    &0.754&0.545 &0.715&0.486 &0.723&0.512  \\
    \hline
    CATHI  &0.752&0.731  &0.710&0.659 &0.756&0.701  &0.820&0.782 \\
    C-ILP  &0.760&0.535 &0.852&0.709 &0.800&0.611  &0.811&0.623 \\
	DeepTrip  &0.765&0.660 &0.831&0.782 &0.834&0.755  &0.808&0.748 \\
	 NASR+  &0.755&0.734 &0.849&0.756 &0.811&0.738  &0.829&0.803\\
	 \hline
	 \hline
	\bf{SelfTrip} &\textbf{0.783}&\textbf{0.779} &\textbf{0.855}&\textbf{0.818}  &\textbf{0.857}&\textbf{0.851}  &\textbf{0.851}&\textbf{0.835} \\
	\bottomrule[1pt]
\end{tabular}
}
\vspace{-0.4cm}
\end{table}
\subsection{Results}
\subsubsection{Overall Performance.}
Table~\ref{E-T} shows the performance comparison of our SelfTrip with several representative benchmarks. Note that we use the bold font to highlight the \textbf{best} performance. Generally, our SelfTrip significantly outperforms all benchmarks across four datasets, with an average improvement of 2.03\%, 6.85\% over the best benchmark with respect to $F_1$ and pairs-$F_1$, respectively. Among the benchmarks including Markov, POIRank and Markov-Rank, the results demonstrate that investigating explicit knowledge, e.g., the human transition regularity and POI co-occurrences, do help in understanding human mobility. However, those methods that either statistically model POI transitions with Markov chain or learn a ranking of POIs with traditional machine learning methods (e.g., RankSVM) cannot automatically explore more complex mobility regularity and handle the data sparsity issue, resulting in an unsatisfactory performance for trip recommendation. In particular, they perform worse than these recent deep representation learning-based methods, such as DeepTrip and NASR+. As for SelfTrip, it performs the best, the plausible reason is that the trip augmentation can address the data sparsity problem and consider the heterogeneity of user demands during POI embedding and trip representation learning. Meanwhile, we investigate the inherent interactions between queries and POIs as the prerequisites for representation learning, which allows us to simulate possible human query demands and trip preference.

\subsubsection{Module Performance.}
\begin{figure}[ht]
\vspace{-0.2cm}
	\centering
	\setlength{\abovecaptionskip}{0pt}
	\setlength{\belowcaptionskip}{0pt}
	\subfigure[POI learning.]{
		\includegraphics[width=0.22\textwidth,height=0.12\textheight]{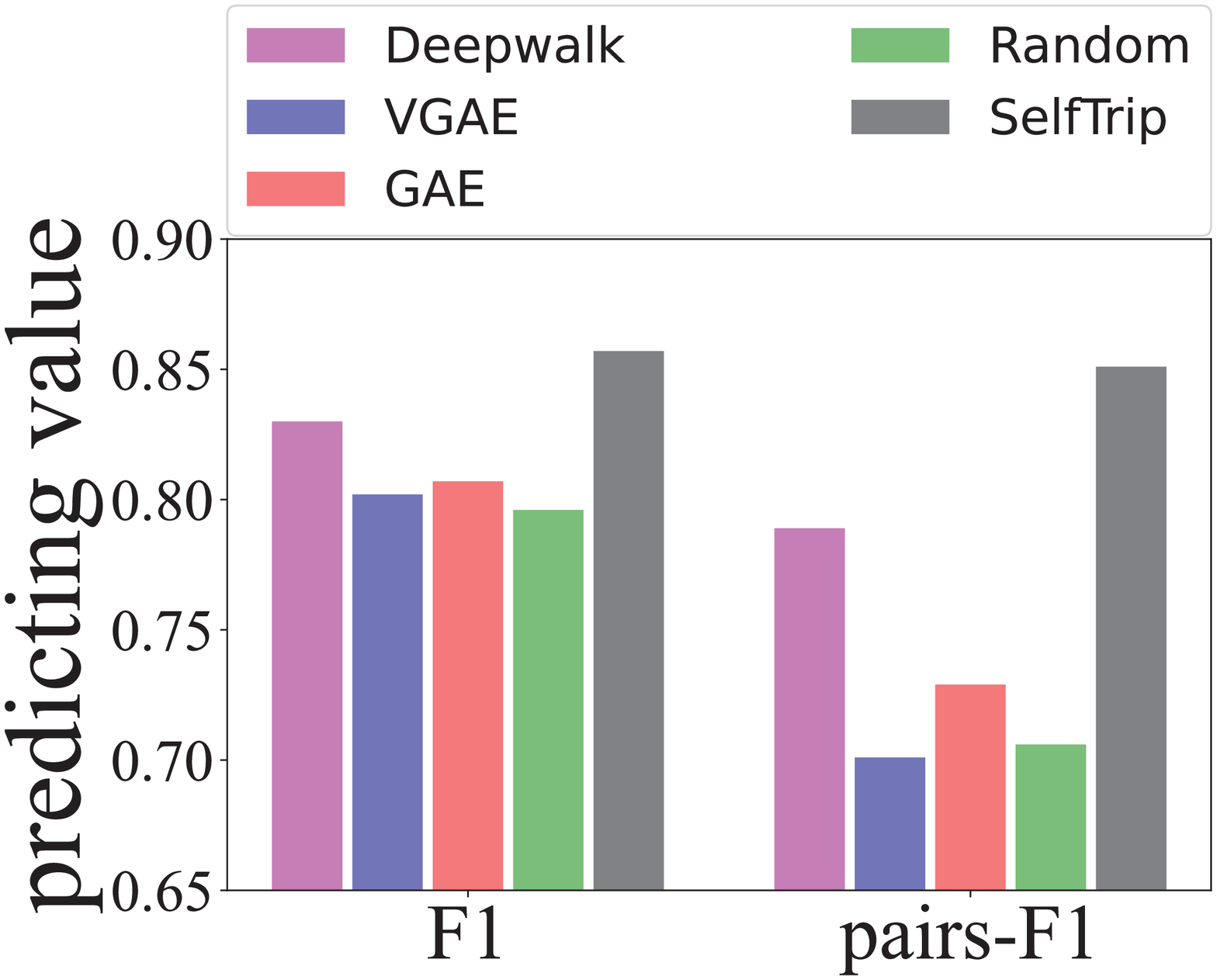}
		\label{V-POI}
	}
	\subfigure[Trip learning.]{
		\includegraphics[width=0.22\textwidth,height=0.12\textheight]{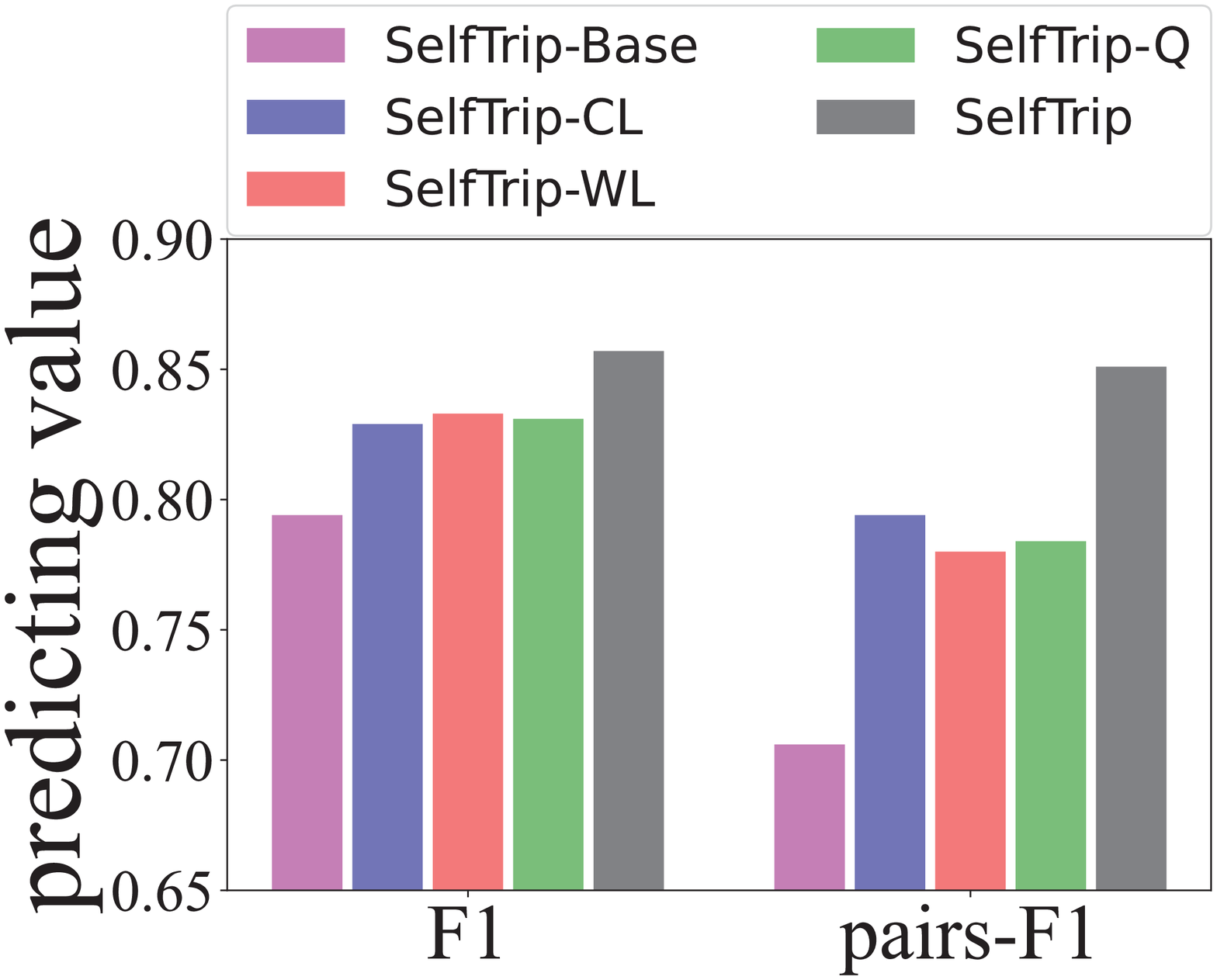}
		\label{V-trip}
	}
	\label{impacts}
	\caption{Ablation study on Glasgow.}
	\vspace{-0.4cm}
\end{figure}

To investigate the capability of POI learning, we compare our self-supervised POI learning with four popular embedding methods, and they are: (1) \textbf{Random} refers to the application of random matrix sampled from Gaussian distribution for POI embeddings; (2) \textbf{Deepwalk}~\cite{perozzi2014deepwalk} uses random walk method to generate the POI sequences on POI graph, and leverages word2vec to formulate the POI embeddings; (3) \textbf{GAE}~\cite{kipf2016variational} utilizes the graph autoencoder to generate node (POI) embeddings; (4) \textbf{VGAE}~\cite{kipf2016variational} is a variant of GAE, integrating the variational Bayes mechanism. As Fig.~\ref{V-POI} shows, we observe that our method significantly outperforms these four embedding methods, and Deepwalk achieves the second best. But Deepwalk is worse than SelfTrip because it only focuses on the surrounding context of a given POI. Note that we observe similar patterns for other datasets and omit to report due to the space limitation. 

To explore the impact of query encoder, contrastive trip learning and the destination-enabled supervised signal. We provide four variants of SelfTrip. The first one is a base model which removes the above three parts and uses a simple concatenation operation for query encoding, called \textbf{SelfTrip-Base}; the second one removes the contrastive learning component in SelfTrip, namely \textbf{SelfTrip-CL}; the third one removes the destination-enabled supervised signal, called \textbf{SelfTrip-WL}, and the last one uses a simple concatenation operation for query encoding, called \textbf{SelfTrip-Q}. As Fig.~\ref{V-trip} shows, SelfTrip performs the best. We find that these three components do help promote performance gains, which further indicates that the sparsity of trip data, the potential destination correlation and query context are important factors in trip recommendation.

\subsubsection{Impact of Augmentation.}
\textit{Query Augmentation.} In this paper, we propose causal random walks to augment the query records. Whether such a method improves the recommendation performance is still a question, especially when the user provides a query demand that never appears before. To answer this question, we re-split each dataset to new training data (about 80\%) and new testing data (about 20\%) where queries in testing data do not appear in the training data. As Table~\ref{query-F1} shows, \textbf{SelfTrip-} denotes without using the augmented queries in SelfTrip. The results demonstrate that query augmentation significantly achieves encouraging gains, especially for the $F_1$ scores regarding Osaka.

\renewcommand{\arraystretch}{0.7} 
\begin{table}[htp]
\vspace{-0.2cm}
	\renewcommand{\multirowsetup}
    {\centering}
	\setlength{\abovecaptionskip}{0pt}
	\setlength{\belowcaptionskip}{0pt}
\caption{Performance comparison.}
\label{query-F1}
\centering
\scriptsize
\setlength{\tabcolsep}{0.8mm}{
\begin{tabular}{l|cc|cc|cc|cc}
	\bottomrule[1pt]
	\multicolumn{1}{l|}{\multirow{2}{*}{\textbf{Method}}}&  
	\multicolumn{2}{c|}{\textbf{Edinburgh}}&\multicolumn{2}{c}{\textbf{Glasgow}}&	\multicolumn{2}{|c|}{\textbf{Osaka}}&\multicolumn{2}{c}{\textbf{Toronto}}\cr
	\cmidrule(lr){2-9}
	\textbf{}  &  $F_1$ & \text{\text{pairs-}}$F_1$& $F_1$ & \text{\text{pairs-}}$F_1$& $F_1$ & \text{\text{pairs-}}$F_1$& $F_1$ & \text{\text{pairs}}$F_1$
	\\
	\hline
	\bf{SelfTrip-}  & 0.593  &\textbf{0.418} &0.729&0.489& 0.669  &0.400 &0.665&\textbf{0.428}  \\
	\hline
	\bf{SelfTrip} &\textbf{0.613} &\text{0.408}  &\textbf{0.755}  &\textbf{0.552}&\textbf{0.694} &\textbf{0.567}  &\textbf{0.677}  &{0.426} \\
	\bottomrule[1pt]
\end{tabular}
}
\vspace{-0.4cm}
\end{table}
\textit{Trip Augmentation.} To evaluate the effect of trip augmentation methods, we empirically show the testing results with different couples of trip augmentation. As Fig~\ref{impacts-data} shows, each cell represents a different couple, where the diagonal means that we only use a single strategy and `None' means we do not use any augmentation methods. We observe that single augmentation method obtains the worse performance while dropout and POI mask are more efficient in trip augmentation. Therefore, we will endeavor to choose more efficient coupling and explore more augmentation methods for data augmentation in the future.
\begin{figure}[ht]
\vspace{-0.4cm}
	\centering
	\setlength{\abovecaptionskip}{0pt}
	\setlength{\belowcaptionskip}{0pt}
	\subfigure[Osaka.]{
		\includegraphics[width=0.22\textwidth,height=0.14\textheight]{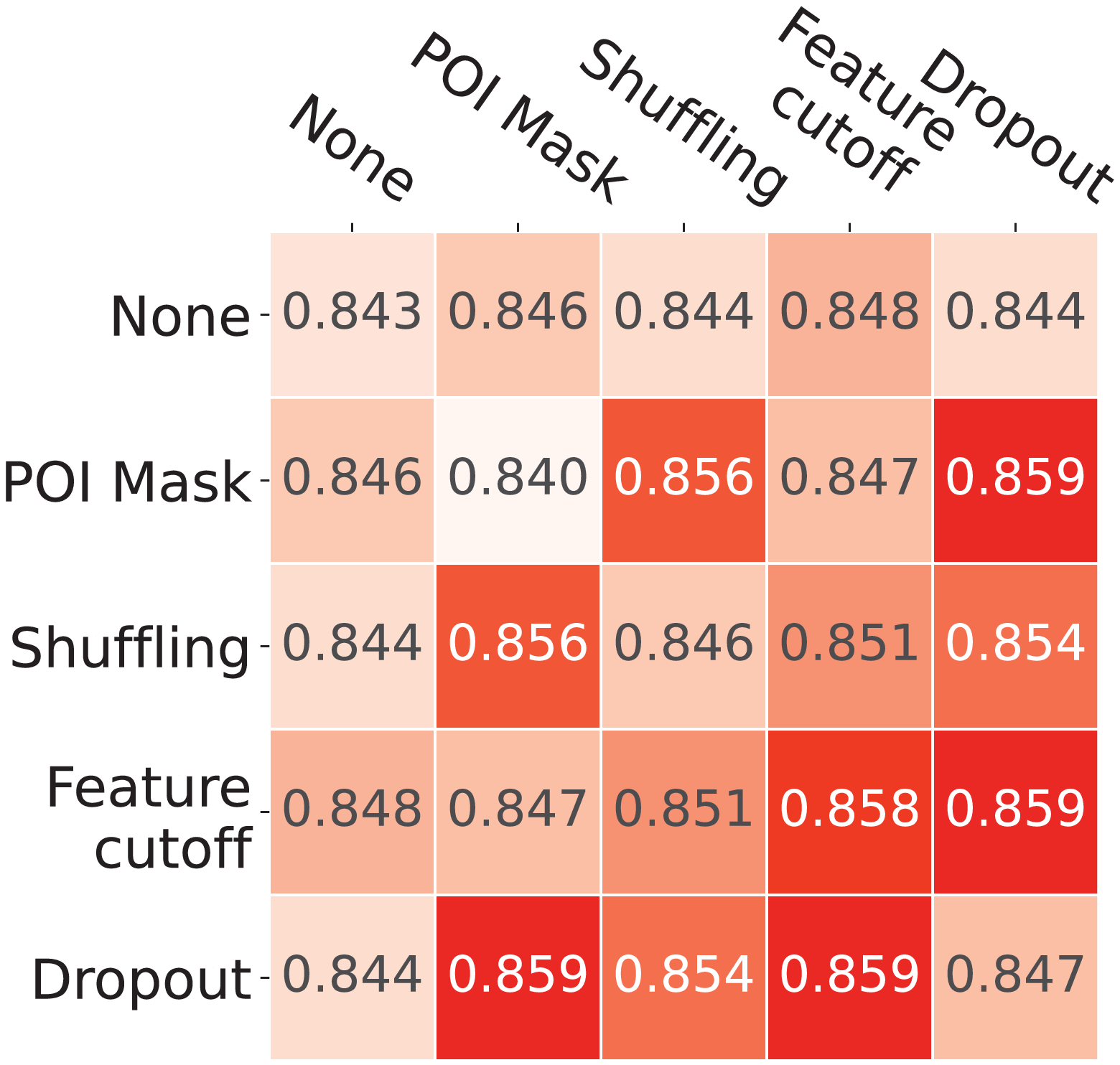}
		\label{V-glas}
	}
	\subfigure[Glasgow.]{
		\includegraphics[width=0.22\textwidth,height=0.14\textheight]{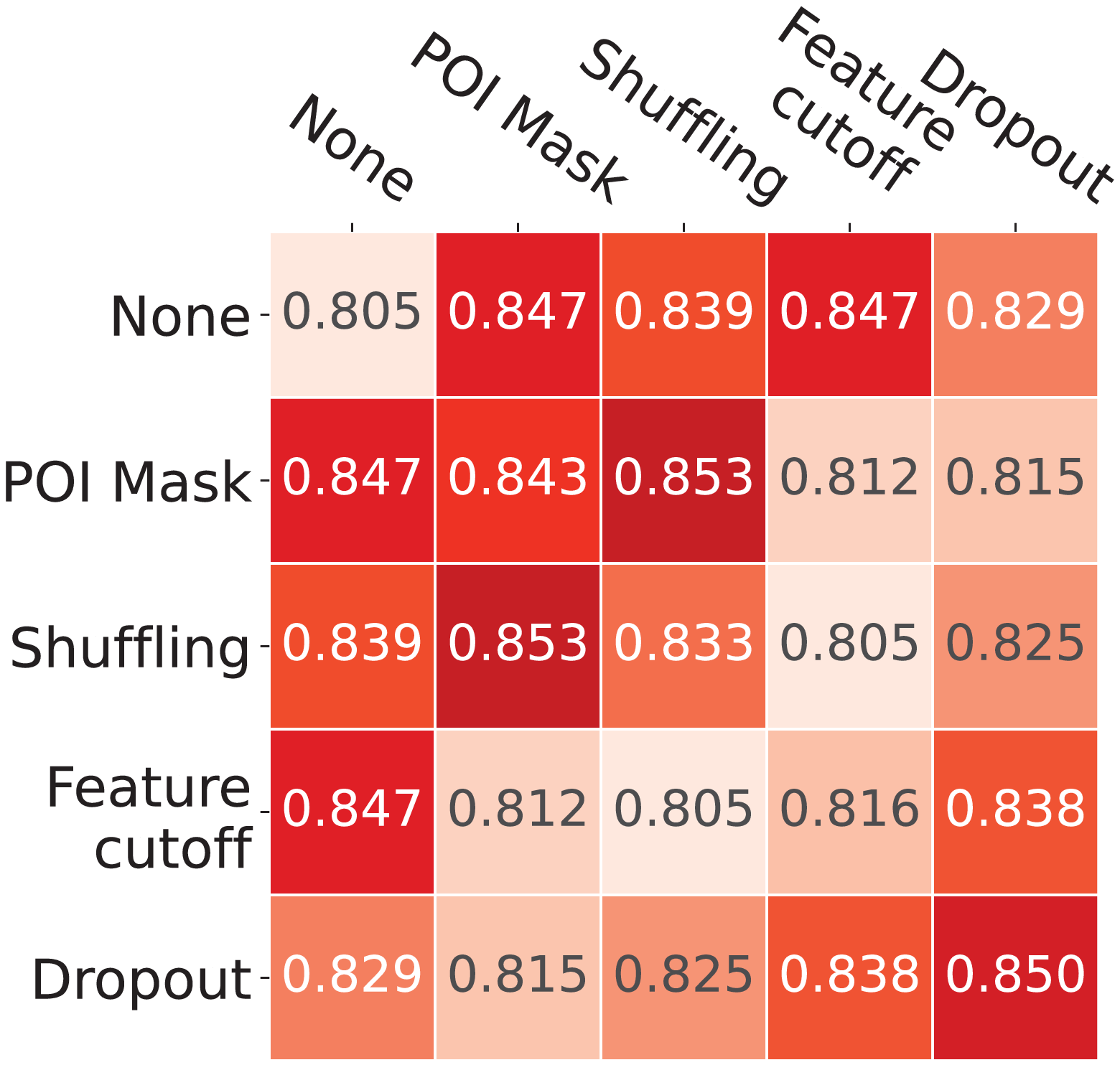}
		\label{V-osaka}
	}
	\caption{$F_1$ score in different trip augmentations}
	\label{impacts-data}
\vspace{-0.5cm}
\end{figure}

\subsubsection{Sensitivity of Hyper-parameters.} 
\begin{figure}[ht]
\vspace{-0.3cm}
	\centering
	\setlength{\abovecaptionskip}{0pt}
	\setlength{\belowcaptionskip}{0pt}
	\subfigure[Hidden size.]{
		\includegraphics[width=0.22\textwidth,height=0.12\textheight]{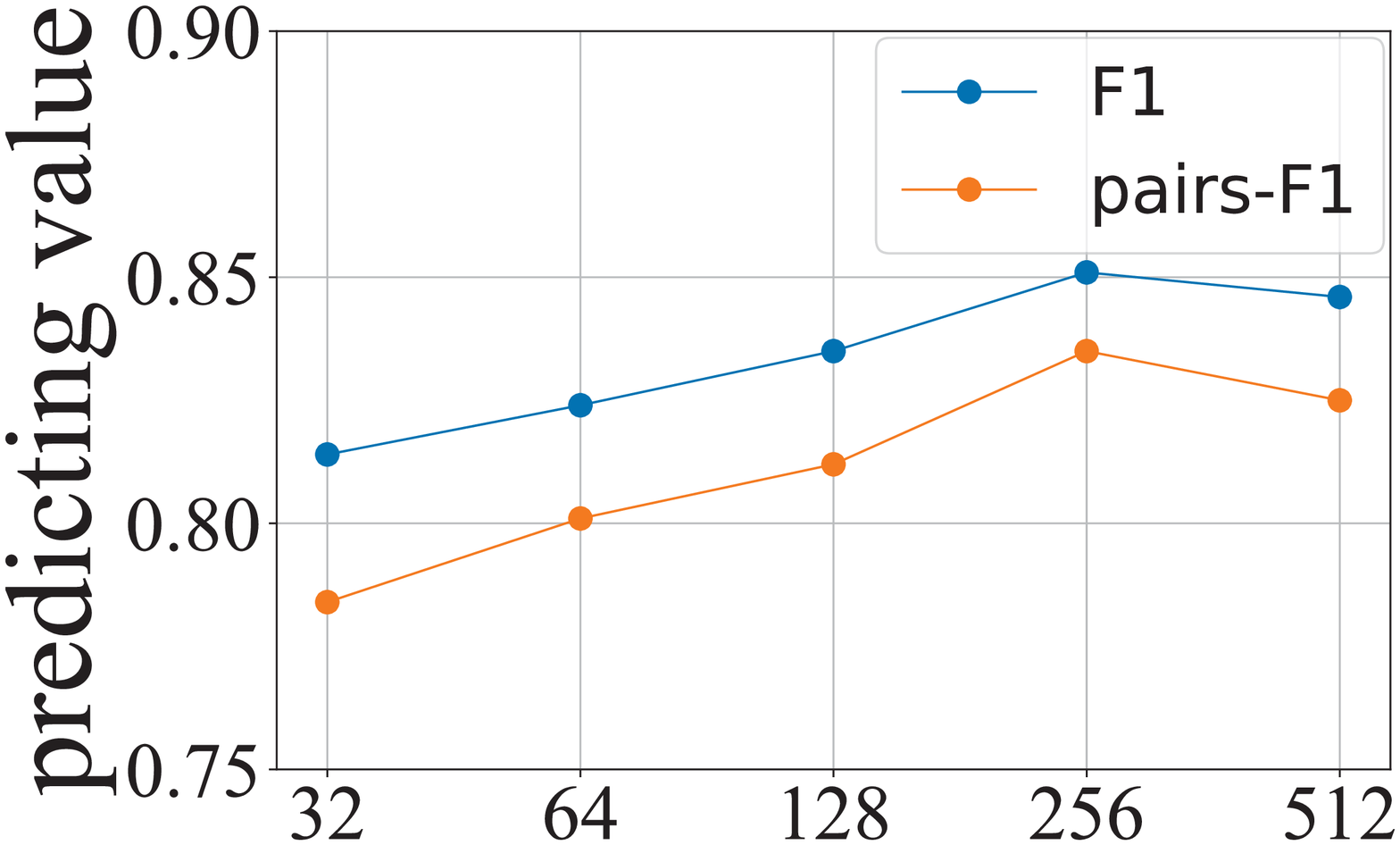}
		\label{h-hidden}
	}
	\subfigure[\#Negative samples]{
		\includegraphics[width=0.22\textwidth,height=0.12\textheight]{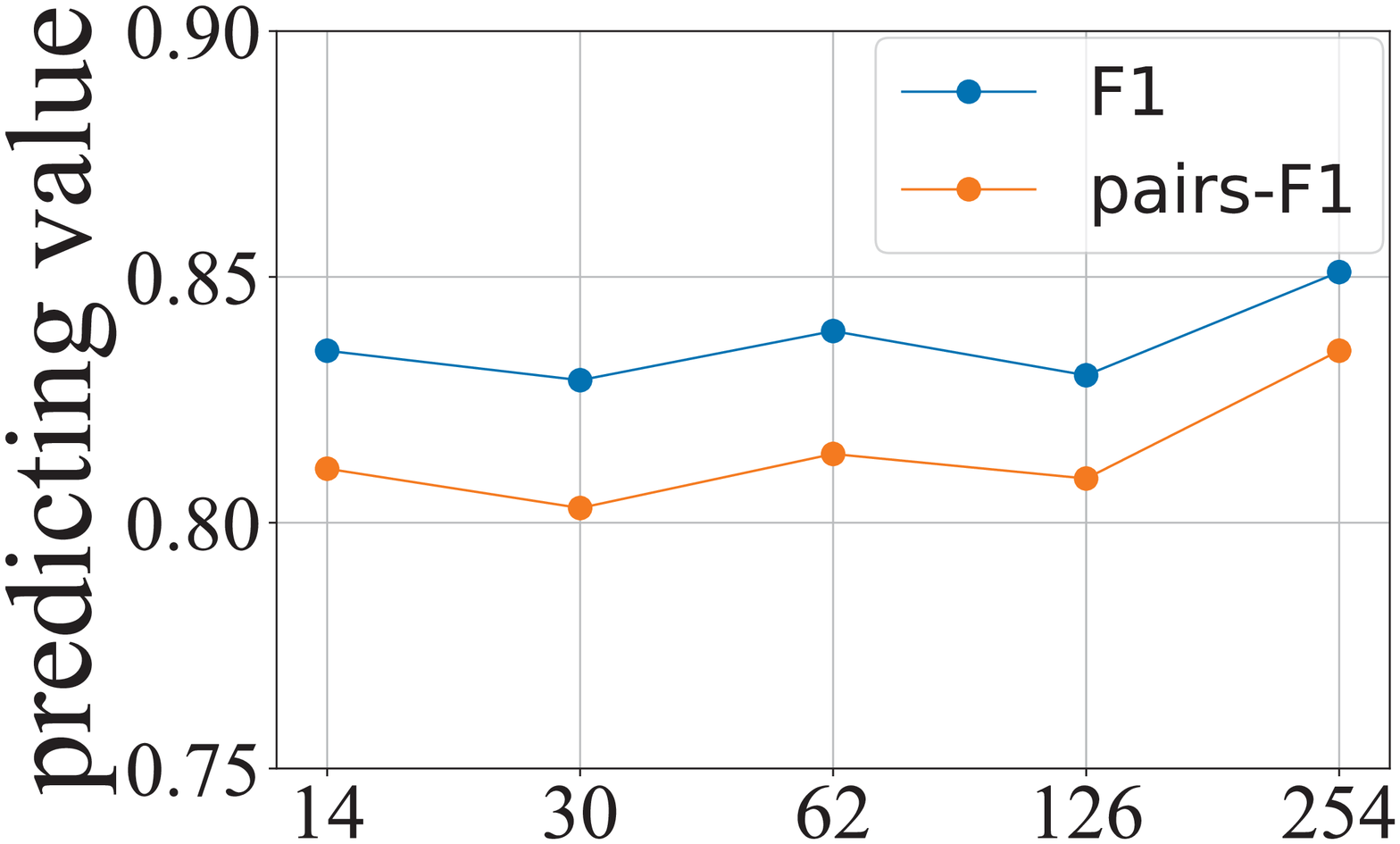}
		\label{h-sample}
	}
	\caption{Hyper-parameter tuning on Toronto.}
	\label{sensitivity}
\vspace{-0.5cm}
\end{figure}

As Fig~\ref{sensitivity} shows, we investigate the sensitivity of some key hyper-parameters, e.g., hidden size and the number of negative samples in trip learning.

\section{Related Work}
\label{Related_Work}
\subsubsection{Trip Recommendation.}
Conventionally, one of the common solutions for trip recommendation is using the simple heuristic models to understand the transitional patterns of different POIs. Chen et al. proposed a probabilistic method by mining the implied information (e.g., transitional matrix) from historical POIs and trips~\cite{chen2016learning}. Gu et al. investigated the attractive routes to make a trip recommendation considering user experience, where a gravity model is leveraged to evaluate the rating score of each attractive route~\cite{gu2020enhancing}. In addition to these, other existing works considered the trip recommendation as a variant of the orienteering problems (OP), maximizing the collected scores from the generated path~\cite{taylor2018travel,he2019joint}. However, they are not good at estimating user diverse demands and historical travel preferences with the limit of data scale, resulting in the inability to learn the real distribution of human travel patterns. Recent researchers used deep representation learning to explore complex relations from POIs and trips, such as semantic proximity of POIs and long-term dependencies in trips~\cite{he2019joint,ho2021user,gao2021adversarial}. He et al. and Ho et al. adopted the popular POI embedding method for trip recommendation, modeling the probability distribution of the co-occurring POIs~\cite{he2019joint,ho2021user}. DeepTrip is a regularized latent variable method to understand human travel patterns, leveraging a trained model for trip recommendation~\cite{gao2021adversarial}. Although the success of deep representation learning, those methods only relied on historical trips, which still suffer the sparsity and insufficient representation issues. In contrast, SelfTrip is a self-supervised framework with data augmentation that can better alleviate the sparsity of queries and trips.

\subsubsection{Self-supervised Representation Learning}

Recent self-supervised learning, which has shown excellent performance in representation learning, brings a new opportunity to address the limit of labeled data scale due to its capability of avoiding the work of
tagging large-scale datasets. And it generally can be categorized into two domains, i.e., generative-based and contrastive-based~\cite{wu2020self}. For example, variational Bayesian and adversarial network attract numerous researchers' attention since they can self-train with the whole data without any label information. Generative-based methods have been widely applied in mobility-based tasks~\cite{liu2020online,wang2019adversarial}, including trip recommendation~\cite{gao2021adversarial}. However, generative-based approaches need to reconstruct accurate local details to learn sample features, while the emerging contrastive learning focuses on implicit multi-views of samples and construct an additional contrastive loss to distill inherent information from the data itself by mutual information maximization principle~\cite{he2020momentum,oord2018representation,xie2021adversarial}. Although the emerging application of contrastive learning has been widely used in CV and NLP, our self-supervised learning framework is among the first work to capture implicit supervised signals from both POI-level and trip-level, which distinguishes from previous studies.

\section{Conclusion and Future Work} \label{sec:concl}
In this paper, we present a systematic self-supervised learning framework to train the POI representation and trip representation for trip recommendation, where a novel model SelfTrip is proposed. We implement a self-supervised learning with data augmentation including query augmentation and trip augmentation to capture the inherent interactions between queries and trips. The empirical results demonstrate that our SelfTrip achieves the best performance. Besides, our proposed self-supervised POI learning significantly outperforms existing widely used embedding methods. In the future, we consider incorporating other features (e.g., social characteristics) to explore human diverse preference. 

\bibliography{bib/CLNext}

\begin{thebibliography}{27}
\providecommand{\natexlab}[1]{#1}

\bibitem[{Aitchison(2021)}]{aitchison2021infonce}
Aitchison, L. 2021.
\newblock InfoNCE is a variational autoencoder.
\newblock \emph{arXiv preprint arXiv:2107.02495}.

\bibitem[{Chen, Ong, and Xie(2016)}]{chen2016learning}
Chen, D.; Ong, C.~S.; and Xie, L. 2016.
\newblock Learning points and routes to recommend trajectories.
\newblock In \emph{Proceedings of the 25th ACM International on Conference on
  Information and Knowledge Management}, 2227--2232.

\bibitem[{Gao et~al.(2017)Gao, Zhou, Zhang, Trajcevski, Luo, and
  Zhang}]{Gao2017}
Gao, Q.; Zhou, F.; Zhang, K.; Trajcevski, G.; Luo, X.; and Zhang, F. 2017.
\newblock Identifying human mobility via trajectory embeddings.
\newblock In \emph{IJCAI}.

\bibitem[{Gao et~al.(2021)Gao, Zhou, Zhang, Zhang, and
  Trajcevski}]{gao2021adversarial}
Gao, Q.; Zhou, F.; Zhang, K.; Zhang, F.; and Trajcevski, G. 2021.
\newblock Adversarial Human Trajectory Learning for Trip Recommendation.
\newblock \emph{IEEE Transactions on Neural Networks and Learning Systems}.

\bibitem[{Gu et~al.(2020)Gu, Song, Jiang, Wang, and Liu}]{gu2020enhancing}
Gu, J.; Song, C.; Jiang, W.; Wang, X.; and Liu, M. 2020.
\newblock Enhancing Personalized Trip Recommendation with Attractive Routes.
\newblock In \emph{Proceedings of the AAAI Conference on Artificial
  Intelligence}, 01, 662--669.

\bibitem[{He, Qi, and Ramamohanarao(2019)}]{he2019joint}
He, J.; Qi, J.; and Ramamohanarao, K. 2019.
\newblock A joint context-aware embedding for trip recommendations.
\newblock In \emph{2019 IEEE 35th International Conference on Data Engineering
  (ICDE)}, 292--303. IEEE.

\bibitem[{He et~al.(2020)He, Fan, Wu, Xie, and Girshick}]{he2020momentum}
He, K.; Fan, H.; Wu, Y.; Xie, S.; and Girshick, R. 2020.
\newblock Momentum contrast for unsupervised visual representation learning.
\newblock In \emph{Proceedings of the IEEE/CVF Conference on Computer Vision
  and Pattern Recognition}, 9729--9738.

\bibitem[{Hinton et~al.(2012)Hinton, Srivastava, Krizhevsky, Sutskever, and
  Salakhutdinov}]{hinton2012improving}
Hinton, G.~E.; Srivastava, N.; Krizhevsky, A.; Sutskever, I.; and
  Salakhutdinov, R.~R. 2012.
\newblock Improving neural networks by preventing co-adaptation of feature
  detectors.
\newblock \emph{arXiv preprint arXiv:1207.0580}.

\bibitem[{Ho and Lim(2021)}]{ho2021user}
Ho, N.~L.; and Lim, K.~H. 2021.
\newblock User Preferential Tour Recommendation Based on POI-Embedding Methods.
\newblock In \emph{26th International Conference on Intelligent User
  Interfaces}, 46--48.

\bibitem[{Kingma and Ba(2014)}]{kingma2014adam}
Kingma, D.~P.; and Ba, J. 2014.
\newblock Adam: A method for stochastic optimization.
\newblock \emph{arXiv preprint arXiv:1412.6980}.

\bibitem[{Kipf and Welling(2016)}]{kipf2016variational}
Kipf, T.~N.; and Welling, M. 2016.
\newblock Variational graph auto-encoders.
\newblock \emph{arXiv preprint arXiv:1611.07308}.

\bibitem[{Lim et~al.(2015)Lim, Chan, Leckie, and Karunasekera}]{Lim2015}
Lim, K.~H.; Chan, J.; Leckie, C.; and Karunasekera, S. 2015.
\newblock Personalized tour recommendation based on user interests and points
  of interest visit durations.
\newblock In \emph{IJCAI}.

\bibitem[{Liu et~al.(2021)Liu, Zhang, Hou, Mian, Wang, Zhang, and
  Tang}]{liu2021self}
Liu, X.; Zhang, F.; Hou, Z.; Mian, L.; Wang, Z.; Zhang, J.; and Tang, J. 2021.
\newblock Self-supervised learning: Generative or contrastive.
\newblock \emph{IEEE Transactions on Knowledge and Data Engineering}.

\bibitem[{Liu et~al.(2020)Liu, Zhao, Cong, and Bao}]{liu2020online}
Liu, Y.; Zhao, K.; Cong, G.; and Bao, Z. 2020.
\newblock Online anomalous trajectory detection with deep generative sequence
  modeling.
\newblock In \emph{2020 IEEE 36th International Conference on Data Engineering
  (ICDE)}, 949--960. IEEE.

\bibitem[{Oord, Li, and Vinyals(2018)}]{oord2018representation}
Oord, A. v.~d.; Li, Y.; and Vinyals, O. 2018.
\newblock Representation learning with contrastive predictive coding.
\newblock \emph{arXiv preprint arXiv:1807.03748}.

\bibitem[{Perozzi, Al-Rfou, and Skiena(2014)}]{perozzi2014deepwalk}
Perozzi, B.; Al-Rfou, R.; and Skiena, S. 2014.
\newblock Deepwalk: Online learning of social representations.
\newblock In \emph{Proceedings of the 20th ACM SIGKDD international conference
  on Knowledge discovery and data mining}, 701--710.

\bibitem[{Taylor, Lim, and Chan(2018)}]{taylor2018travel}
Taylor, K.; Lim, K.~H.; and Chan, J. 2018.
\newblock Travel itinerary recommendations with must-see points-of-interest.
\newblock In \emph{Companion Proceedings of the The Web Conference 2018},
  1198--1205.

\bibitem[{Wang, Wu, and Zhao(2021)}]{wang2021personalized}
Wang, J.; Wu, N.; and Zhao, X. 2021.
\newblock Personalized Route Recommendation with Neural Network Enhanced A*
  Search Algorithm.
\newblock \emph{IEEE Transactions on Knowledge and Data Engineering}.

\bibitem[{Wang et~al.(2019)Wang, Fu, Xiong, and Li}]{wang2019adversarial}
Wang, P.; Fu, Y.; Xiong, H.; and Li, X. 2019.
\newblock Adversarial substructured representation learning for mobile user
  profiling.
\newblock In \emph{Proceedings of the 25th ACM SIGKDD International Conference
  on Knowledge Discovery \& Data Mining}, 130--138.

\bibitem[{Wu et~al.(2020)Wu, Wang, Feng, He, Chen, Lian, and Xie}]{wu2020self}
Wu, J.; Wang, X.; Feng, F.; He, X.; Chen, L.; Lian, J.; and Xie, X. 2020.
\newblock Self-supervised Graph Learning for Recommendation.
\newblock \emph{arXiv preprint arXiv:2010.10783}.

\bibitem[{Wu et~al.(2019)Wu, Lian, Jin, and Chen}]{wu2019graph}
Wu, Y.; Lian, D.; Jin, S.; and Chen, E. 2019.
\newblock Graph Convolutional Networks on User Mobility Heterogeneous Graphs
  for Social Relationship Inference.
\newblock In \emph{IJCAI}, 3898--3904.

\bibitem[{Xie et~al.(2021)Xie, Liu, Zhang, Lu, Wang, and
  Ding}]{xie2021adversarial}
Xie, Z.; Liu, C.; Zhang, Y.; Lu, H.; Wang, D.; and Ding, Y. 2021.
\newblock Adversarial and Contrastive Variational Autoencoder for Sequential
  Recommendation.
\newblock In \emph{Proceedings of the Web Conference 2021}, 449--459.

\bibitem[{Yan et~al.(2021)Yan, Li, Wang, Zhang, Wu, and Xu}]{yan2021consert}
Yan, Y.; Li, R.; Wang, S.; Zhang, F.; Wu, W.; and Xu, W. 2021.
\newblock ConSERT: A Contrastive Framework for Self-Supervised Sentence
  Representation Transfer.
\newblock \emph{arXiv preprint arXiv:2105.11741}.

\bibitem[{Yu et~al.(2020)Yu, Cui, Guo, Lu, Li, and Lu}]{yu2020category}
Yu, F.; Cui, L.; Guo, W.; Lu, X.; Li, Q.; and Lu, H. 2020.
\newblock A category-aware deep model for successive poi recommendation on
  sparse check-in data.
\newblock In \emph{Proceedings of the web conference 2020}, 1264--1274.

\bibitem[{Zhou et~al.(2020{\natexlab{a}})Zhou, Wu, Trajcevski, Khokhar, and
  Zhang}]{zhou2020semi}
Zhou, F.; Wu, H.; Trajcevski, G.; Khokhar, A.; and Zhang, K.
  2020{\natexlab{a}}.
\newblock Semi-supervised Trajectory Understanding with POI Attention for
  End-to-End Trip Recommendation.
\newblock \emph{ACM Transactions on Spatial Algorithms and Systems (TSAS)},
  6(2): 1--25.

\bibitem[{Zhou et~al.(2019)Zhou, Yue, Trajcevski, Zhong, and
  Zhang}]{zhou2019context}
Zhou, F.; Yue, X.; Trajcevski, G.; Zhong, T.; and Zhang, K. 2019.
\newblock Context-aware variational trajectory encoding and human mobility
  inference.
\newblock In \emph{The World Wide Web Conference}, 3469--3475.

\bibitem[{Zhou et~al.(2020{\natexlab{b}})Zhou, Wang, Zhao, Zhu, Wang, Zhang,
  Wang, and Wen}]{zhou2020s3}
Zhou, K.; Wang, H.; Zhao, W.~X.; Zhu, Y.; Wang, S.; Zhang, F.; Wang, Z.; and
  Wen, J.-R. 2020{\natexlab{b}}.
\newblock S3-rec: Self-supervised learning for sequential recommendation with
  mutual information maximization.
\newblock In \emph{Proceedings of the 29th ACM International Conference on
  Information \& Knowledge Management}, 1893--1902.

\end{thebibliography}
\end{document}